\documentstyle[12pt,MERLpub]{article}

\title{Renormalization Group Approach To Error-Correcting Codes}
\author{Jonathan S. Yedidia\thanks{* MERL, 201 Broadway, Cambridge MA 02139. 
{\sf yedidia@merl.com}} and Jean-Philippe Bouchaud\thanks{' SPEC CEA-Saclay, Orme des Merisiers, 91191 Gif Sur Yvette, France. {\sf bouchaud@spec.saclay.cea.fr}}} 
\pubnumber{2001-19}
\pubabstract{We explain an algorithm that approximately but
efficiently assesses particular 
parity-check error-correcting codes of large, but
finite, blocklength. 
This algorithm is based on the ``renormalization-group'' 
approach from physics: the idea is to continually replace an error-correcting
code with a simpler error-correcting code that has 
nearly identical performance,
until the code is reduced to a small enough size that its performance
can be computed exactly.
This assessment algorithm can be used as a subroutine in a more general
algorithm to search for optimal error-correcting codes of
specified blocklength and rate.}
\pubhistory{1. May 2001 Draft version \\
2. June 2001 Current version}
\def\BE{\begin{equation}}
\def\EE{\end{equation}}
\def\BEA{\begin{eqnarray}}
\def\EEA{\end{eqnarray}}
\input{psfig}
\begin{document}
\MERLreport
\pagestyle{plain}
\section{Introduction}
A fundamental problem in the field of information theory is the design of
optimal or nearly optimal 
error-correcting codes of given block-length and rate that can 
also be decoded
practically. This problem
can now be considered essentially solved in the small blocklength (e.g. 
$N<100$)
and very large blocklength (e.g. $N > 10^6$) regimes. However, error-correcting
codes that are used in practical situations (for example, for wireless
communication) typically have blocklengths in an intermediate regime (around
$N = 2000$). (The reason that intermediate blocklength codes are used in 
practice is that larger blocklength codes have better
performance, but have a longer decoding time, so one will normally
choose the code with the
largest blocklength for which the 'lag' caused by decoding is
still tolerable.)

In the small
blocklength regime, classical coding theory,
as summarized in textbooks such as 
\cite{MacWilliams77}, provides a panoply
of codes of different blocklengths and rates,
many of which are known to be optimal or nearly optimal. As long
as the blocklength is small enough, these codes can be
also be decoded practically (and optimally) using maximum-likelihood decoders. 

In the last few years,
the problem of finding good codes in the very large blocklength regime has
been essentially solved but in a very different way; by focusing 
on parity-check codes defined using sparse (generalized)
parity check matrices \cite{MackaySparse}. These
kinds of codes were first introduced by Gallager in 1962 \cite{Gallager63}, 
but were not 
properly appreciated until recently. In the last eight years, however,
new and 
improved codes defined by sparse generalized parity check matrices
(such as turbocodes \cite{Berrou93, Mackay99}, 
irregular low-density parity check (LDPC)
codes \cite{Luby97, Luby2001, Davey99, Richardson2001, Chung2001}, Kanter-Saad
codes \cite{KanterSaad99, KanterSaad2000, MackaySparse},
repeat-accumulate codes \cite{RAcodes}, and irregular repeat-accumulate codes \cite{IRAcodes}) have
been the object of intense study.
Such codes have
three particularly noteworthy advantages. First, 
they can be efficiently decoded using belief propagation (BP) 
iterative decoding \cite{MackayEtAl97}.
Secondly, their performance can often be 
theoretically analyzed in the infinite-blocklength limit using 
the {\it density evolution} approach \cite{RichardsonUrbanke2001}.
Finally, using the density evolution approach, or through simulations,
one can demonstrate that these
codes are good codes,
in the sense that in the infinite-blocklength
limit, BP decoding will perfectly decode all message blocks that have a noise
level below some threshold level, and that threshold level 
is often not too far from the
Shannon limit.

In recent years, a favored way to design new codes has thus been to
optimize codes for the infinite blocklength limit using density evolution, and to
hope that a scaled-down version would still be a good code \cite{Luby2001, Richardson2001, Chung2001, IRAcodes}.
The problem with this approach is that for $N < 10^4$ at least, we are
still noticeably far from the infinite-blocklength limit. In particular, 
simulations will
find many decoding failures at noise levels far below the threshold level
predicted by infinite blocklength calculations. Furthermore, there will not
necessarily
even exist a way to scale down the codes derived from the density evolution
approach. For example, the
best known irregular LDPC codes at a given rate (in the $N \rightarrow \infty$
limit) will often have variable nodes 
that should participate in hundreds or even thousands of parity
checks \cite{Chung2001}, 
which obviously makes no sense if the overall number of
parity checks is $100$ or less. When one wants to make
real codes of finite blocklength, one is therefore 
often forced to choose a code
which is sub-optimal in the infinite-blocklength limit, with no theoretical
guidance.

Our goal, which is achieved by the renormalization group
 approach described here, has 
therefore been to develop an assessment algorithm more powerful than the
ordinary density evolution approach, which will predict, at least 
approximately,
the decoding failure rate
as a function of the noise level for a specific 
code of finite blocklength. It is 
important to realize that for finite blocklengths, one does not expect
perfect decoding below any particular 
threshold noise level, so that to evaluate
a code, one now needs a whole performance curve rather than a single
number (the critical noise threshold) 
as might be computed in the density evolution approach.

The outline of the rest of this paper is as follows. In the next section,
we review the density evolution approach for the binary erasure channel,
where it is particularly simple. We pay particular attention to codes
defined on trees, for which the density evolution approach becomes exact. 
Section 3 is the heart of the paper, where 
we introduce and explain our renormalization-group (RG) approach. We show
how it recovers exact answers for codes defined on trees, and give a procedure,
which can be made increasingly accurate at the cost of more computational
power to handle codes defined on graphs with loops. We present
some numerical results comparing our RG calculations 
with simulations of realistic
finite blocklength codes in 
section 4. In section 5, we explain how to extend the RG approach to the
Additive White Gaussian Noise (AWGN) channel. In
section 6, we speculate on how one might use our RG algorithm as a 
sub-routine for a more general algorithm for the design of codes.

\section{The density evolution approach for the binary erasure channel (BEC)}
The density evolution approach is analytically very simple for the
binary erasure channel. \cite{Luby97, Bazzi99} Since this approach is important background for our 
own RG
approach, we will review it in this section. 

\subsection{Parity check codes}
We will begin by studying linear block binary codes
which can be defined in terms of an ordinary
parity check matrix. 
In a parity check matrix $A$, the columns
represent transmitted variable 
bits, while the rows define linear constraints between
the variable bits. More specifically, the matrix $A$ defines a set of
valid vectors (codewords) $z$, such that each component of $z$ is 0 or 1,
and 
\BE
A z = 0
\EE
where we assume all multiplication and addition is modulo 2.

If a parity check matrix has
$N$ columns and $N-k$ rows it will represent a code of blocklength
$N$ and rate $k/N$ (unless some of the rows are not linearly independent, in
which case some of the parity checks are redundant, and the
code will actually be of higher rate).

\begin{figure}[thb]
\centerline{
	\psfig{file=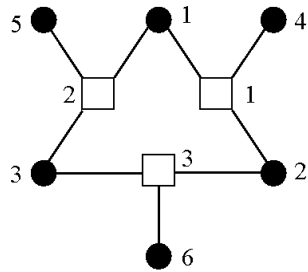,height=3in}}
\caption[]{Tanner graph for a simple error-correcting code.}
\label{fig:tanner6}
\end{figure}

For each parity check matrix, there is a corresponding Tanner graph
representation. \cite{Tanner81} 
A Tanner graph is a bipartite graph with two kinds of nodes:
variable nodes (which we denote by circles) and check nodes (denoted by
squares).
In a Tanner graph, each check node is connected
to the variable nodes that represent the bits involved in that check.
For example, the parity check matrix
\BE
A = \left( \begin{array}{cccccc}
1 & 1 & 0 & 1 & 0 & 0 \\
1 & 0 & 1 & 0 & 1 & 0 \\
0 & 1 & 1 & 0 & 0 & 1 \end{array}
\right)
\EE
corresponds to the Tanner graph shown in figure \ref{fig:tanner6}.

Codes represented by parity check matrices are ``linear,'' which means that
all the codewords are linear combinations of other codewords. There will be 
$2^k$ codewords, each of length $N$; e.g., 
for the example given above, the codewords are
{\tt 000000}, {\tt 001011}, {\tt 010110}, {\tt 011101}, {\tt 100110},
{\tt 101101}, {\tt 110011}, {\tt 111000}. Because of the linearity property,
we may use any of the codewords as a representative; throughout this
paper, we will always assume that the all-zeros codeword is transmitted.

\subsection {Belief propagation decoding in the BEC}
The binary erasure channel is a binary input channel with three output symbols:
a {\tt 0}, a {\tt 1}, and an erasure, which can be represented by 
a question mark {\tt ?}. The input symbol will pass through the
channel as an erasure with
probability $x$ and will be
received correctly with probability $1-x$. It is
important to note 
that the BEC never flips bits from {\tt 0} to {\tt 1} or vice versa.
If we assume that the all-zeros codeword is transmitted, all received
words will thus consist entirely of zeros and erasures. 

We will assume that the receiver decodes using a
belief propagation (BP) decoder with discrete messages. 
A message $m_{ia}$
will be sent from each
variable node $i$ to each check $a$ that it participates in, with the message
representing information about the state of the variable node. In general,
the message
can be in one of three states: {\tt 1}, {\tt 0}, or {\tt ?}, but since  
we assume that the all-zeros state is always transmitted,
we can ignore the possibility that $m_{ia}$ has value {\tt 1}.

Similarly, there will be a message $m_{ai}$ sent from each check node
$a$ to all
the variable nodes $i$ that participate in that check. These messages should
be interpreted as directives from the check to the variable node about what
state it should be in, based on the states of the other variable nodes 
participating in the check.
The check-to-bit messages can again in principle
take on the values {\tt 0}, {\tt 1}, or {\tt ?}, 
but again only the two messages {\tt 0} and {\tt ?} will be relevant when the
all-zeros codeword is transmitted.

In the BP decoding algorithm for the BEC,
a message $m_{ia}$ from a variable node to a check node will be equal to
a non-erasure received message (because such messages are always correct in
the BEC),
or to an erasure if all incoming messages are erasures.
A message $m_{ai}$ from a check node $a$ to a variable node $i$ 
will be an erasure if any incoming message
from another node participating in the check is an erasure, otherwise
it will take on the value of the binary sum of all incoming messages from other
nodes participating in the check.

The BP decoding algorithm is an iterative algorithm. One should initialize the
messages so that all variable nodes that are not erased send out messages 
equal to the corresponding received bit, and all other messages are initially 
erasures. Iterating the BP message equations, one will eventually always
converge to stationary messages (convergence of BP decoding algorithms is
guaranteed for the particularly simple 
BEC, but not for other channels). The final decoded value
of any erased variable node is just the value of any non-erasure message
coming into that node, unless there is no incoming non-erasure message, in
which case the BP decoding algorithm gives up and fails to decode that
particular variable node.

\subsection{Density evolution}
We now consider the average of BP decoding over many blocks. 
Associated with each message
$m_{ia}$, we introduce 
a real number $p_{ia}$ which represents the probability
that the message $m_{ia}$ is an erasure. Similarly, we
associate with each message $m_{ai}$ a real number $q_{ai}$ which
represents the probability that the message $m_{ai}$ is an erasure.

In the density evolution approach, we compute the probabilities $p_{ia}$ and
$q_{ai}$ in a way that is exact as long as the Tanner graph representing the
code has no loops. We take
\BE
\label{p_de_general}
p_{ia} = x \prod_{b \in N(i) \backslash a} q_{bi}
\EE
where $b \in N(i) \backslash a$ represents all check nodes that neighbor
variable node $i$ except for check node $a$. This equation can be derived
from the fact
that for a message $m_{ia}$ to be an erasure, the variable node $i$
must be erased in transmission, and all incoming messages from other checks
must be erasures as well. Of course, if the incoming messages were 
correlated, this equation would not be correct, but on a Tanner graph with
no loops, each incoming message is independent of the others.

Similarly, we find that 
\BE
\label{q_de_general}
q_{ai} = 1 - \prod_{j \in N(a) \backslash i} (1-p_{ja})
\EE
which can be derived (again assuming incoming messages are uncorrelated) 
from the fact that a message $q_{ai}$ will only be in 
a {\tt 0} or {\tt 1} state if all incoming messages are in a {\tt 0} or
{\tt 1} state.

The density evolution 
equations (\ref{p_de_general}) and (\ref{q_de_general}) can be solved by
iteration. A good initialization is $p_{ia} = x $ for all messages from
variable nodes to check nodes and $q_{ai} = 0$ for all messages from check
nodes to variable nodes, as long as one begins the iteration with the $q_{ai}$
messages. The BEC density evolution equations should ultimately
converge (this can be guaranteed for codes defined on graphs without loops). 
One can finally compute $b_i$, which
is the probability of a failure to decode at variable node $i$, from the
formula
\BE
\label{b_de_general}
b_i = x \prod_{a \in N(i)} q_{ai}
\EE

\subsection{Exact solution of a small code}
\begin{figure}[thb]
\centerline{
	\psfig{file=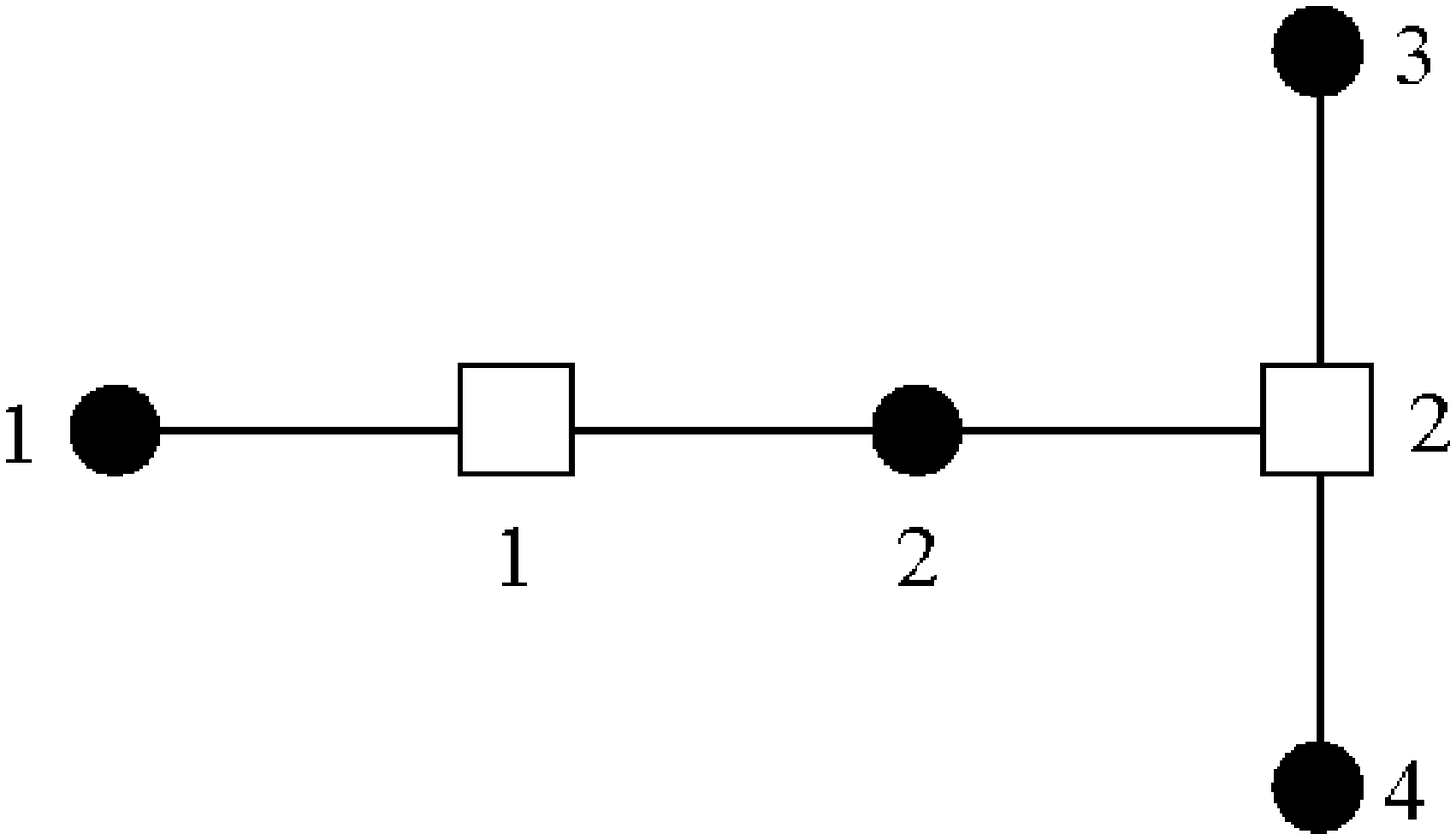,height=3in}}
\caption[]{}
\label{fig:tanner4}
\end{figure}

As mentioned, the density evolution equations (\ref{p_de_general}),
(\ref{q_de_general}), and (\ref{b_de_general}) 
should be exact when the code has a Tanner graph
representation without loops. Let us work through a small example, to
see how this works. We consider the code with parity check matrix
\BE
A = \left( \begin{array}{cccc}
1 & 1 & 0 & 0 \\
0 & 1 & 1 & 1 \end{array}
\right)
\EE
and a corresponding Tanner graph shown in figure \ref{fig:tanner4}. 

This code has
four codewords: {\tt 0000}, {\tt 0011}, {\tt 1101}, and {\tt 1110}. If
the {\tt 0000} message is transmitted, there will be
sixteen possible received messages: {\tt 0000}, {\tt 000?}, {\tt 00?0},
{\tt 00??}, {\tt 0?00}, and so on. The probability of receiving a particular
message with $n_e$ erasures is $x^{n_e} (1-x)^{(4-n_e)}$. Messages might
be partially or completely decoded by a BP decoder; for example the
received message {\tt ?00?} will be fully decoded to {\tt 0000}, but the
message {\tt 0???} will only be partially decoded to {\tt 00??}, because there
is not enough information to determine whether the transmitted codeword
was actually {\tt 0000} or {\tt 0011}. 

We can easily compute the exact
probability that a given bit will remain an erasure after decoding by summing
over the sixteen possible received messages weighted by their probabilities.
For example, the first bit will only 
be decoded as an erasure if one of the following
messages are received: {\tt ???0}, {\tt ??0?}, or {\tt ????}, so the total
probability that the first bit will not be decoded is 
$2 x^3 (1-x) + x^4 = 2 x^3 - x^4$.
If we focus on the last bit instead, we find that it will be decoded unless
one of the following messages is sent: {\tt 00??}, {\tt 0???}, {\tt ?0??},
{\tt ??0?} or {\tt ????}, so the overall probability that the fourth bit
is not decoded will be 
$x^2 (1-x)^2 + 3 x^3 (1-x) + x^4 = x^2 + x^3 - x^4$.

In the density evolution approach, we need to solve equations
for the following
variables: $p_{11}$, $p_{21}$, $p_{22}$, $p_{32}$, $p_{42}$, $q_{11}$,
$q_{12}$, $q_{22}$, $q_{23}$, $q_{24}$, $b_{1}$, $b_2$, $b_3$, and $b_4$. 
The equations are:
\BEA
p_{11} = x \\
p_{21} = x q_{22} \\
p_{22} = x q_{12} \\
p_{32} = x \\
p_{42} = x \\
q_{11} = p_{21} \\
q_{12} = p_{11} \\
q_{22} = 1 - (1-p_{32})(1-p_{42}) \\
q_{23} = 1 - (1-p_{22})(1-p_{42}) \\
q_{24} = 1 - (1-p_{22})(1-p_{32})
\EEA
and
\BEA
b_1 = x q_{11} \\
b_2 = x q_{12} q_{22} \\
b_3 = x q_{23} \\
b_4 = x q_{24}
\EEA
Solving these equations, we find
\BEA
p_{11} = x \\
p_{21} = 2 x^2 - x^3 \\
p_{22} = x^2 \\
p_{32} = x \\
p_{42} = x \\
q_{11} = 2 x^2 - x^3 \\
q_{12} = x \\
q_{22} = 2 x - x^2 \\
q_{23} = x + x^2 - x^3 \\
q_{24} = x + x^2 - x^3 
\EEA
and
\BEA
b_{1} = 2 x^3 - x^4 \\
b_{2} = 2 x^3 - x^4 \\
b_{3} = x^2 + x^3 - x^4 \\
b_{4} = x^2 + x^3 - x^4.
\EEA
Examining the results for $b_1$ and $b_4$, we see that the density evolution
solution agrees exactly with the direct approach for this code.

\subsection{The large blocklength limit}
If we assume that all local neighborhoods look identical, we can simplify
the density evolution equations. 
For example, if each variable node belongs to $d_v$ parity
checks, and each check node is attached to $d_c$ variable nodes, then we
can take all the $p_{ia}$ equal to the same value $p$, all the $q_{ai}$
equal to the same value $q$, and all $b_i$ equal to the
same value $b$. We then find
\BE
\label{eq_pde_reg}
p = x q^{d_v-1} 
\EE
\BE
\label{eq_qde_reg}
q = 1 - (1-p)^{d_c-1}
\EE
and
\BE
b = x q^{d_v}
\EE
 which are the density evolution equations for $(d_v,d_c)$ {\it regular
Gallager codes}, valid in the $N \rightarrow \infty$ limit. A regular Gallager
code \cite{Gallager63} is a code defined by a sparse random 
parity check matrix characterized by the restriction that
each row has exactly $d_c$ 1's in it, and each column contains 
exactly $d_v$ 1's.
The intuitive reason that
these equations are valid in the infinite blocklength limit is that
as $N \rightarrow \infty$, the size of typical loops in the Tanner graph of
a regular Gallager code will also go to infinity, so all incoming messages
to a node will be independent, and a regular Gallager code
will behave like a code defined on a graph without loops.

If we solve equations (\ref{eq_pde_reg}) and 
(\ref{eq_qde_reg}) for specific values
of $d_v$ and $d_c$, we find that below a critical erasure threshold $x_c$,
the solution is $p=q=b=0$, 
which means that decoding is perfect. Above $x_c$, $b$ will
have a non-zero solution, which correspond to decoding failures.
$x_c$ is easy to determine numerically. For example, 
if $d_v=3$ and $d_c=5$, then $x_c \approx 0.51757$.

These density evolution calculations can be generalized to irregular
Gallager codes \cite{Luby97}, or other 
codes like irregular repeat-accumulate codes \cite{IRAcodes} which
have a finite number of different classes of nodes with different
neighborhoods. In this generalization, one derives a system of equations,
typically with one equation for the messages leaving each class
of node. By solving the system of equations, one can again find a critical
threshold $x_c$, below which decoding is perfect. 
Such codes can thus be optimized in the $N \rightarrow \infty$ limit
by finding the code that has maximal noise threshold $x_c$.
Simulations of such codes with very large blocklengths agree quite
well with the density evolution predictions.

Unfortunately, the density evolution approach is useless, or at least
misleading, for codes with
finite blocklength. One might think that one could solve equations
(\ref{p_de_general}) and (\ref{q_de_general}) 
for any finite code, and hope that
ignoring the presence of loops one is not too important a mistake.
This does not work out, as one can simply see by considering regular Gallager
codes. Equations (\ref{p_de_general}), (\ref{q_de_general}), 
and (\ref{b_de_general})
for
a finite blocklength regular Gallager code will have exactly the same solutions
as one would find in 
the infinite-blocklength limit, so one would not predict {\it any} finite-size
effects. Simulations, on the other hand, show that
the real performance of finite-blocklength regular Gallager codes is
considerably different (and worse) 
than that predicted by such a naive approach.

\section{The renormalization group approach}
\subsection{Intuition}
The basic idea behind the ``real-space'' renormalization group approach 
from physics \cite{RSRGbook}
is very similar to the
idea behind recursion from computer science. To evaluate the performance
of a large but finite code, we try to replace the code with a slightly
smaller code with the same performance. In particular, at each step in the
process, we keep a Tanner graph and a set of $p_{ia}$ and $q_{ai}$ variables 
just as in the density evolution approach. We will call the 
combination of a Tanner graph 
and the $p$ and $q$ variables a ``decorated Tanner graph.'' 
The heart of the RG approach
is the RG transformation, by which we eliminate (``renormalize away'') one
node in the decorated Tanner graph, and adjust the remaining values of
the $p$ and $q$ messages so that the new code has a decoding failure rate
as close as possible to the old code. With each renormalization step, the
decorated Tanner graph representing our code will thus shrink by one
node, until it is finally
small enough that the performance of the code can be computed exactly in
an efficient way.

We will explain all the details in the following sub-sections, but in 
general, the RG algorithm will work as follows:
\begin{enumerate}
\item Choose a ``target'' variable 
node $i$ for which we want to compute the decoding
failure rate $b_i$.
\item While the number of nodes remaining in the graph is greater than 
the number that one can comfortably handle exactly, 
repeatedly renormalize away nodes 
from the graph according to the following procedure:
  \begin{enumerate}
  \item Mark the ``distance'' of every
node from the ``target'' node. The
distance between two nodes is the minimal number of nodes that one
needs to pass through on the graph to travel from one node to the other.
  \item As long as 
there are any ``leaf'' (a {\it leaf} is a node which is only
connected to one other node in the graph) check or variable nodes, 
renormalize them away. The order in which they are renormalized away will
not matter, but for concreteness, we will renormalize those furthest from
the ``target'' node first, breaking ties randomly.
  \item Otherwise, choose a single 
variable node from among those furthest from the
target node, that has the fewest neighboring check nodes, and renormalize it
away.
  \end{enumerate}
\item Compute $b_i$ for the remaining graph exactly.
\end{enumerate} 

It should be clearly understood that the RG approach is approximate, 
and that there exists considerable freedom in the implementation.
Different choices made in the implementation will lead to slightly
different results. One can deal with this problem by constructing a series of
systematically
better RG approximations that should eventually converge to the exact answer.
We shall see how this works out in our case in section 
\ref{subsection:loops}.
In the physics literature, this type of approach is well-known; the
interested reader should consult the book \cite{RSRGbook}.

\subsection{The RG transformation for Tanner graphs with no loops}
First we consider loop-free Tanner graphs, and write down the RG 
transformations that are sufficient to
give exact results for such codes.
In later subsections, we will extend the RG
transformations in order to obtain good approximate
results for Tanner graphs with loops.

We will always
initialize our decorated Tanner graph such that all $b_i=x$, $p_{ia}=x$ and all
$q_{ai} = 0$. Imagine that we are interested in the decoding failure rate
$b_i$ at a specific node $i$. Our procedure will be to
obtain $b_i$ by repeatedly renormalizing
away nodes, other than the variable node $i$ itself, 
that are ``leaves'' of the decorated Tanner 
graph.

The first possibility that we need to concern ourselves with is when we
renormalize away a ``leaf'' 
variable node $i$ that is connected to a single check node
$a$. Clearly, when the node $i$ vanishes, $p_{ia}$ and $q_{ai}$ will also be
discarded. We need to renormalize all the $q_{aj}$ variables leading out
of the check $a$ to other nodes $j$. Our formula will be
\BE
\label{rg_trans_q}
q_{aj} \leftarrow 1 - (1-q_{aj})(1-p_{ia})
\EE
where the left arrow indicates that we replace the old value of $q_{aj}$ with
this new value. Notice that each renormalization of $q_{aj}$ will increase its
value. 

When we renormalize away a ``leaf'' 
check node $a$ that is only connected to a single
variable node $i$, we need to adjust the values of all the $p_{ib}$ variables
leading to other checks $b$ that node $i$ is attached to. The renormalization
group transformation will be
\BE
\label{rg_trans_p}
p_{ib} \leftarrow p_{ib} q_{ai}.
\EE
Notice that each renormalization of $p_{ib}$ will decrease its value.
At the same time, we should also renormalize the $b_i$ as follows:
\BE
\label{rg_trans_b}
b_i \leftarrow b_i q_{ai}.
\EE
When only the ``target'' node $i$ remains, we can just read off the current
value of $b_i$ and that will serve as the RG prediction.

\subsection{A small example}
The RG procedure might be easier to understand if we work through a small
example. Recall the code defined by the parity check matrix
\BE
A = \left( \begin{array}{cccc}
1 & 1 & 0 & 0 \\
0 & 1 & 1 & 1 \end{array}
\right).
\EE

\begin{figure}[ht!]
\centerline{
	\psfig{file=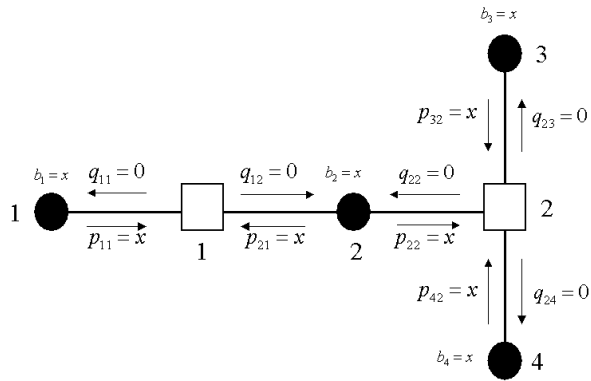,height=3in}}
\caption[]{Decorated Tanner graph}
\label{fig:3a}
\end{figure}

\begin{figure}[ht!]
\centerline{
	\psfig{file=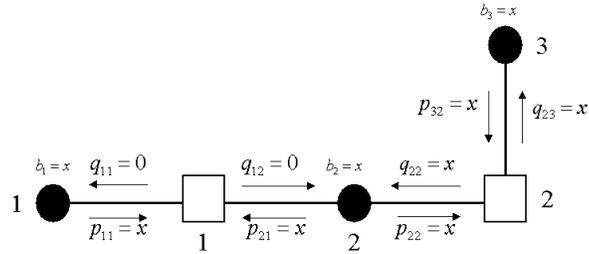,height=3in}}
\caption[]{Decorated Tanner graph after renormalizing variable node 4.}
\label{fig:3b}
\end{figure}

\begin{figure}[ht!]
\centerline{
	\psfig{file=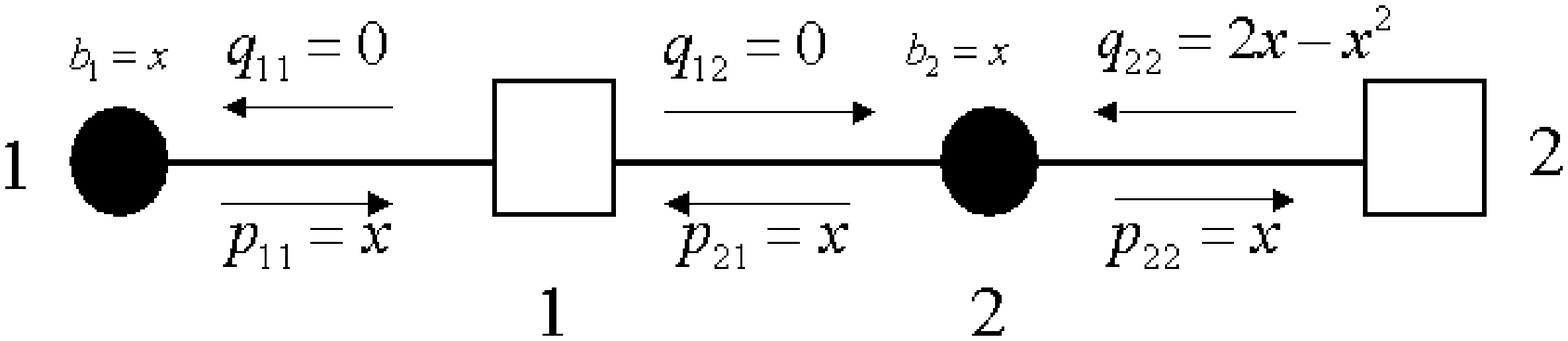,height=3in}}
\caption[]{Decorated Tanner graph after renormalizing variable nodes 3 and 4.}
\label{fig:3c}
\end{figure}

\begin{figure}[ht!]
\centerline{
	\psfig{file=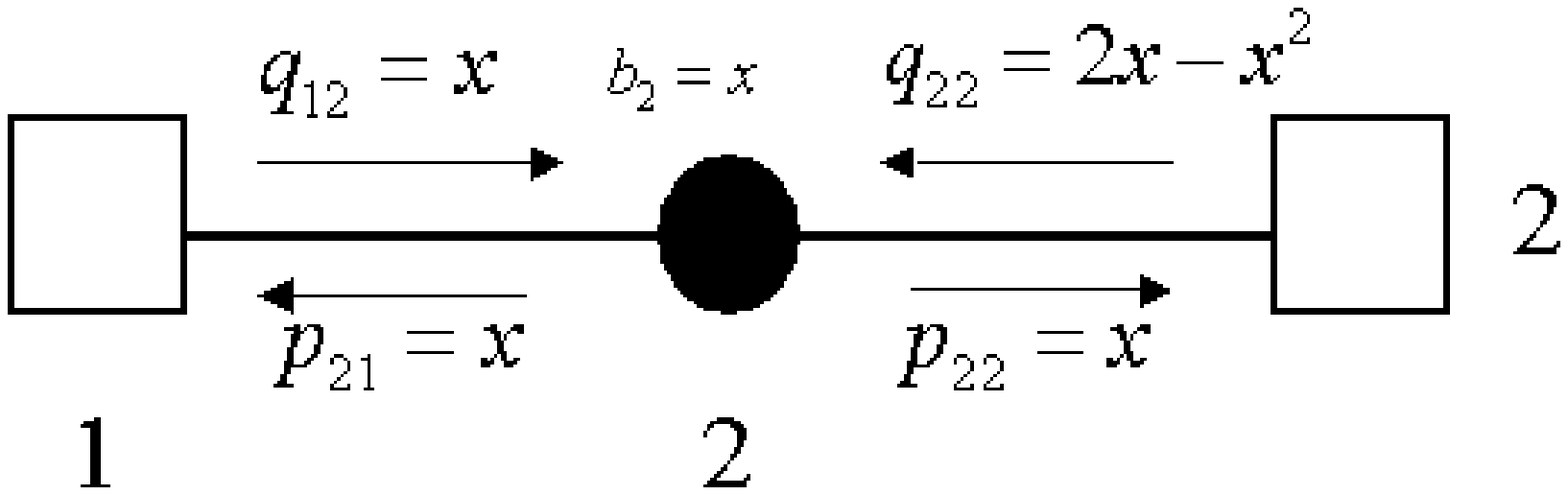,height=3in}}
\caption[]{Decorated Tanner graph after renormalizing variable nodes 1, 3 and 4.}
\label{fig:3d}
\end{figure}

\begin{figure}[th!]
\centerline{
	\psfig{file=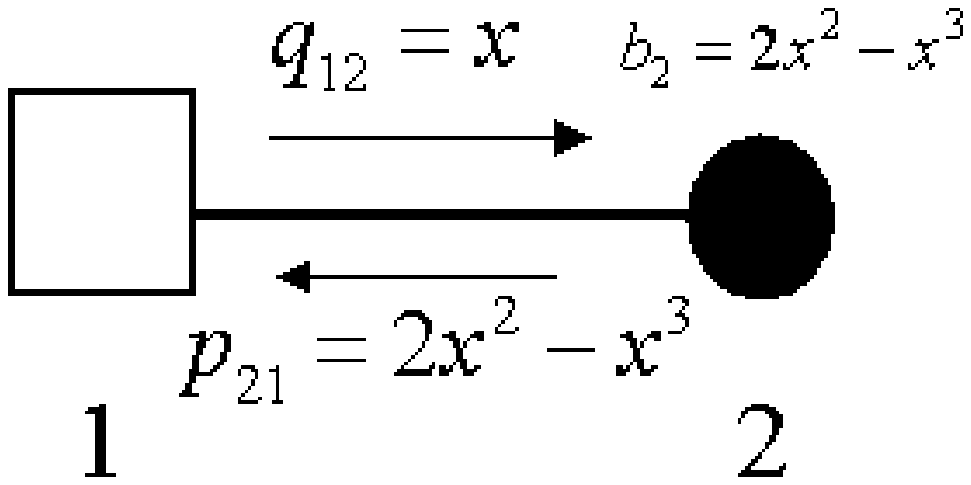,height=3in}}
\caption[]{Decorated Tanner graph after renormalizing variable nodes 1, 3 and 4, and check node 2.}
\label{fig:3e}
\end{figure}

Let us imagine that we would like to compute the decoding failure rate at
the second variable node $b_2$. We initialize 
$p_{11}=p_{21}=p_{22}=p_{32}=p_{42}=x$, $q_{11}=q_{12}=q_{22}=q_{23}=q_{24}=0$,
and $b_2=0$. In figure \ref{fig:3a}, we show the decorated Tanner graph for this
code. All of the variable nodes other than variable 
node $2$ are leaf nodes, so we 
can renormalize any of them away. According to our general algorithm, we
should renormalize away the one furthest from node $2$, breaking ties 
randomly. Let's say we choose variable
 node $4$. Then we discard $p_{42}$ and
$q_{24}$ and obtain new values $q_{22}=x$ and $q_{23}=x$ using 
equation (\ref{rg_trans_q}). The new decorated Tanner graph is shown in 
figure \ref{fig:3b}. Let's next renormalize away variable node $3$. We discard $p_{32}$
and $q_{23}$ and renormalize $q_{22}$ to the value $1-(1-x)^2 = 2 x - x^2$.
The new decorated Tanner graph is shown in figure \ref{fig:3c}. 
Next we renormalize
away variable node 1. 
We discard $p_{11}$ and $q_{11}$ and obtain the new renormalized
value $q_{12}=x$; the Tanner graph is now shown if figure \ref{fig:3d}. 
Next we
renormalize away check node $2$. We can discard $p_{22}$ and $q_{22}$ and
obtain $p_{21} = b_2 = 2 x^2 - x^3$ (shown in figure \ref{fig:3e}.) 
Finally we
renormalize away check node $1$. We are left with only a single node
(our original node $2$) and $b_2$ gets renormalized to its correct value
$b_2 = 2 x^3 - x^4$. 

This example makes it clear why the RG approach is exact for a code
defined on a graph without loops: the RG transformations essentially
reconstruct the density evolution equations, and we know that
density evolution is exact for such codes. As we shall see, the advantage
of the RG approach is that it still gives a good approximation for codes
defined on graphs with loops.

\subsection{The RG approach for a graph with loops}
\label{subsection:loops}
\begin{figure}[htb]
\centerline{
	\psfig{file=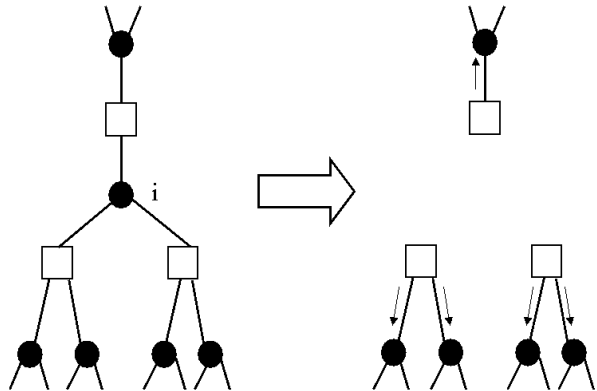,height=3in}}
\caption[]{Removing the non-leaf node $i$. The arrows indicate the $q$ 
variables that will be renormalized as a result.}
\label{fig:break_loops}
\end{figure}

For a code defined a graph that has loops, we will eventually have to 
renormalize away a variable node $i$ that is not a ``leaf'' node. (Note 
that we could also renormalize away non-leaf check nodes by defining the
appropriate RG transformations, but we will 
choose instead to always renormalize
away non-leaf variable nodes.)
To do 
that, we first 
collect all the check nodes $a$, $b$, etc., that node $i$ is attached to.
Obviously, we will discard $q_{ai}$, $q_{bi}$, $p_{ia}$, $p_{ib}$, etc.
For any given check node attached to $i$ (say check node $a$), we must 
also collect all
the other variable nodes $j$ attached to $a$, and renormalize the values of
$q_{aj}$. In figure \ref{fig:break_loops}, we illustrate the process of
removing a non-leaf node. 

 The renormalization of the $q_{aj}$ variable 
can be done to varying degrees of
accuracy. The simplest approach would be to use equation (\ref{rg_trans_q})
directly.
The problem with this approach is that the value of $p_{ia}$ which is used
will always be an over-estimate. Recall that $p_{ia}$ decreases with every
renormalization. Since we are renormalizing away the $i$th node before it
has become 
a leaf node, $p_{ia}$ has not yet been fully renormalized, and is thus
over-estimated.

Instead of using $p_{ia}$ directly, we could use the value that it would have
after we renormalized away all the checks connected to it; that is we
could replace $p_{ia}$ in equation (\ref{rg_trans_q}) with an effective
$p_{ia}^{\rm eff}$ given by
\BE
p_{ia}^{\rm eff} = p_{ia} \prod_{b \in N(i) \backslash a} q_{bi}.
\EE
On the other hand, we know that
the values of the $q_{bi}$ are {\it under-estimates} since they have not
yet been fully renormalized either, so $p_{ia}^{\rm eff}$ as written above
would also be an under-estimate.
We could 
attempt to correct this mistake by going further another level: before we 
estimate a $p_{ia}^{\rm eff}$, we first re-estimate 
the $q_{bi}$ which feed into it. Thus, we replace
the $p_{ia}$ in equation (\ref{rg_trans_q}) with an effective 
$p_{ia}^{\rm eff}$ given by
\BE
p_{ia}^{\rm eff} = p_{ia} \prod_{b \in N(i) \backslash a} q_{bi}^{\rm eff},
\EE
where $q_{bi}^{\rm eff}$ is in turn given by
\BE
q_{bi}^{\rm eff} = 1 - (1-q_{bi}) \prod_{k \in N(b) \backslash i} (1-p_{kb}).
\EE
Putting all these together, we finally get the RG transformation
\BE
\label{rg_trans_q_noleaf}
q_{aj} \leftarrow 1 - (1-q_{aj}) \left(1-p_{ia} \prod_{b \in N(i) \backslash a}
\left[1-(1-q_{bi})\prod_{k \in N(b) \backslash i}(1-p_{kb})\right]\right)
\EE

\begin{figure}[htb]
\centerline{
	\psfig{file=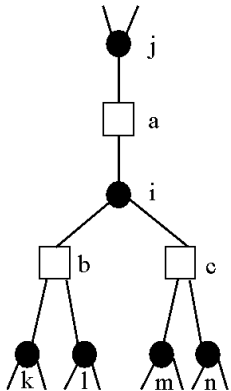,height=3in}}
\caption[]{Node $i$ sees a local tree-like structure.}
\label{fig:local_tree}
\end{figure}

The RG transformation (\ref{rg_trans_q_noleaf}) 
is worth explaining in more detail.
In figure \ref{fig:local_tree}, we illustrate the equation for a case where variable node $i$ is
attached to three checks node $a$, $b$, and $c$, and check 
node $a$ is in turn attached
to a variable node $j$. Check nodes $b$ and $c$ in turn are connected to 
their own variable nodes labeled $k$, $l$, $m$, and $n$. 
We would like to know the new probability $q_{aj}$
that check
node $a$ will send variable node $j$ an erasure message, 
taking into account the information that flows through node $i$.
We already 
have some previous accumulated probability $q_{aj}$ that check node $a$ sends
variable node $j$ 
an erasure message (because of 
other nodes previously attached to $a$ that have already been
renormalized). The new probability of an erasure message can be figured out
from a logical argument: 
``$m_{aj}$ will be an erasure it was already, {\bf or} 
if $m_{ia}$ is an erasure {\bf and}
($m_{bi}$ {\bf or} $m_{kb}$ {\bf or} $m_{lb}$ are erasures) {\bf and} 
($m_{ci}$ {\bf or} 
$m_{mc}$ {\bf or} $m_{nc}$ are erasures).''
Converting such a logical argument into an equation for probabilities is 
straightforward: when we see ``$m_1$ {\bf and} $m_2$'' for two 
statistically independent 
messages in 
a logical argument, it translates to $(p_1 p_2)$ for the corresponding 
probabilities, while ``$m_1$ {\bf or} $m_2$'' translates to 
$(1 - (1-p_1) (1-p_2))$. Converting our full logical argument for figure
4 into an
equation for probabilities, we thus recover an example 
of the RG transformation 
(\ref{rg_trans_q_noleaf}).

\begin{figure}[htb]
\centerline{
	\psfig{file=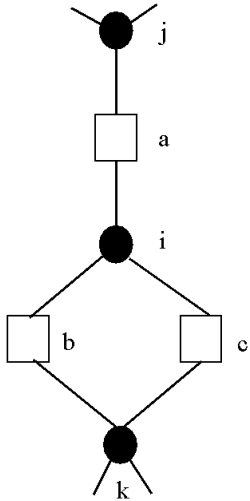,height=3in}}
\caption[]{Node $i$ sees a local neighborhood with loops.}
\label{fig:local_loop}
\end{figure}

We should always take our RG transformation for $q_{aj}$ to correspond to
the logic of the local neighborhood around the node $i$ that we are removing.
In fact, the RG transformation given in equation (\ref{rg_trans_q_noleaf})
is only 
appropriate if the local neighborhood of node $i$ is tree-like, and must be
corrected if there are short loops in the local neighborhood.
For example, in figure \ref{fig:local_loop}, we illustrate a case where a variable 
node $k$ is attached
to two check nodes $b$ and $c$ which are each attached to the node $i$ that
we plan
to remove. First consider the renormalization of $q_{aj}$. 
Note that before check nodes $b$ or $c$ are renormalized,
the probabilities $p_{kb}$ and $p_{kc}$ that variable node $k$ sends out an
erasure must be identical, because all renormalizations of $p_{kb}$ and
$p_{kc}$ happen in tandem. Our logic argument for whether check node
$a$ will send variable node $j$ an erasure message would thus be:
``$m_{aj}$ will be an erasure if it was already, {\bf or} if ($m_{ia}$ is
an erasure) {\bf and} (($m_{kb}$ is an erasure) {\bf or} ($m_{bi}$
{\bf and} $m_{ci}$ are erasures)).'' (We have used the fact that at this stage
in the renormalization process, if $m_{kb}$ is
an erasure, $m_{kc}$ must be as well.) Converting our logic argument into
an RG transformation, we get
\BE
q_{aj} \leftarrow 1-(1-q_{aj})(1-p_{ia}(1-(1-p_{kb})(1- q_{bi} q_{ci})))
\EE

The appropriate renormalizations of $q_{bk}$ and $q_{ck}$ are more
complicated: the messages $m_{bk}$ and $m_{ck}$ are correlated
because of node $i$, and we must keep track of that correlation
after node $i$ is removed. We have tried several relatively ad-hoc
rules for assigning renormalized values to $q$'s that all arrive at the same
node (such as renormalizing the product $q_{bk} q_{ck}$ as a whole), 
but found the results to be unsatisfactory because they depended
sensitively on the details of the rules. In general, to 
correctly account for the correlations caused by such short loops, 
we shall need to introduce additional variables beyond the $q$ and $p$
variables that we use here.  We defer a detailed discussion of this complex
issue to another paper \cite{YBinprep}. In this paper, we will restrict our
examples to
codes where the local structure is always tree-like and 
such short loops do not exist.

The procedure we are describing for renormalizing a non-leaf variable node
$i$
can be made increasingly accurate by increasing the size of the neighborhood
around the node $i$ that is treated correctly. Naturally, as we increase
the size of the neighborhood, we must pay for the increased accuracy with
greater computation. We will use the following terminology: if, when 
renormalizing the node $i$, we use the values of $p_{ia}$ directly, we
will say that the resulting RG transformations have ``depth'' of one. If
we first adjust the values of $p_{ia}$ by considering all the check nodes 
$a$
attached to $i$ and all the variable nodes $k$ 
attached to those check nodes,
we will say the resulting RG transformations (e.g. those described above)
have a depth of two. If we go one step further and also consider
the check nodes attached to the variable nodes $k$ and the variable nodes
attached to those check nodes, we say the RG transformations have a
depth of three, and so on.

\subsection{Finishing the RG computation exactly}
In the RG approach, we can always renormalize nodes away until we are left
with just our ``target'' node $i$, and then read off the decoding failure
rate for that node $b_i$. On the other hand, after we have renormalized away
enough nodes, we could just as well finish the computation exactly. 

For the purposes of describing the exact computation, we assume that we
are given a Tanner graph of $N$ nodes, and associated with each node $i$ is
an erasure probability 
$x_i$. (This is a little different from the decorated Tanner
graph we are used to dealing with, but we shall show how to convert a 
Tanner graph into such a form.) 
To exactly compute the decoding failure rate of a given node $i$,
we generate all $2^N$ possible received 
message blocks (ranging from the
correct all-zeros message all the way to the all-erasures message), and decode
each of them using a BP decoder. Each message block has a probability
\BE
p = \prod x_i \prod (1-x_j)
\EE
where the first product is over all nodes that are erased and the second
product is over all nodes that are not erased. 
We simply compute $b_i$ by taking the
weighted average over all received messages 
of the probability that node $i$ decodes to an erasure. 
Of course, the complexity
of the exact calculation is $O(2^N)$, so we are restricted to small $N$, but
nevertheless one can gain some accuracy by switching to an exact calculation
after one has renormalized away enough nodes.

\begin{figure}[htb]
\centerline{
	\psfig{file=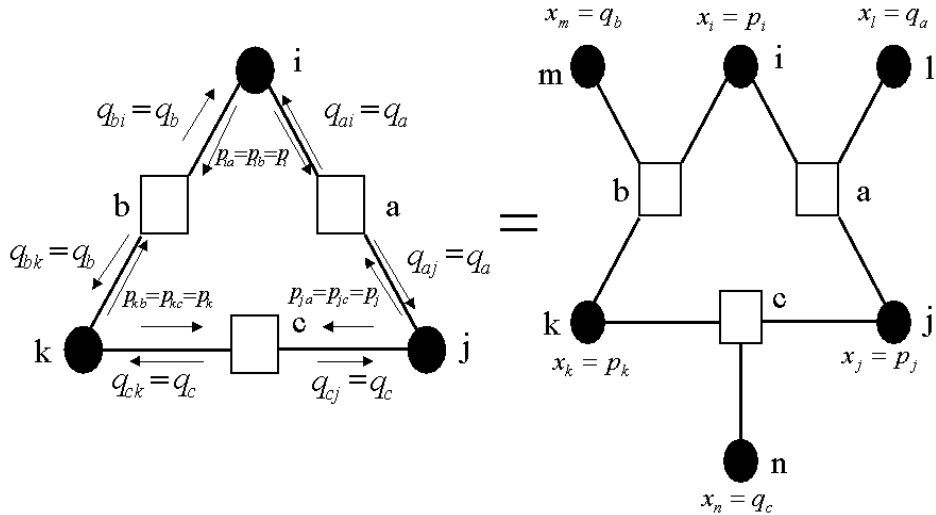,height=4in}}
\caption[]{Expanding a decorated Tanner graph into an equivalent Tanner
graph with erasure probabilities.}
\label{fig:6}
\end{figure}

The one subtlety in the exact final calculation is that one needs a
Tanner graph and the associated erasure probabilities 
at each node, but in the RG
approach, we
manipulate decorated Tanner graphs. Fortunately, it is easy to convert
a decorated Tanner graph into the appropriate form. Note that at each 
step of the RG approach, all the probabilities $q_{ai}$ leading out of
the check node $a$ must be equal (we say $q_{ai} = q_a$)
and all the probabilities $p_{ia}$ leading
out of the variable node $i$ will be equal (we say $p_{ia} = p_i$). 
We can set all the $q_{a}$ 
probabilities equal
to zero if we expand the graph by adding 
a new variable node $k$ to node $a$ with $p_{ka}=q_a$.
When we are left with a decorated Tanner graph such that
all $q$ probabilities are zero, and all $p_{ia}$ 
probabilities coming out of each variable
node are equal to $p_i$, we may interpret the $p_i$ as the erasure 
probabilities of the variable nodes. In figure \ref{fig:6}, we give an example of
expanding a decorated Tanner graph into an equivalent Tanner graph with
erasure probabilities.

\subsection{Extension to generalized parity check matrices}
\begin{figure}[htb]
\centerline{
	\psfig{file=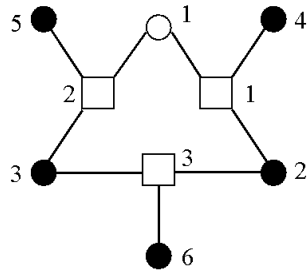,height=3in}}
\caption[]{A Wiberg graph.}
\label{fig:7}
\end{figure}

Many of the best modern codes, such as turbo-codes, Kanter-Saad codes, and
repeat-accumulate codes, are easily represented in terms of 
{\it generalized} parity check matrices \cite{MackaySparse}. In a generalized
parity check matrix, additional columns are added to a parity check
matrix which represent ``hidden nodes''--state variables which are not
transmitted. A good notation for the state variables is a horizontal line
above the corresponding columns. For example, we would write
\BE
\label{gpc}
A = \left( \begin{array}{cccccc}
\overline{1} & 1 & 0 & 1 & 0 & 0 \\
1 & 0 & 1 & 0 & 1 & 0 \\
0 & 1 & 1 & 0 & 0 & 1 \end{array}
\right)
\EE
to indicate a code where the first variable node was a hidden node. To
indicate that a variable node is a hidden node in our graphical model, we
use an open circle rather than a filled-in circle. Such a graph, which 
generalizes Tanner graphs, is called a ``Wiberg graph'' 
\cite{Wiberg96, Wiberg95}. In figure \ref{fig:7}, 
we give the Wiberg graph corresponding
to the code defined by the generalized parity check matrix (\ref{gpc}).

The generalization of our RG procedure to handle Wiberg graphs is very
straightforward. We initialize the probabilities $p_{ia}$ coming out
of a hidden node at 1, instead of at the erasure rate $x$ as we do for
ordinary transmitted variable nodes. This reflects the fact that hidden
nodes are automatically erased, while ordinary variable nodes are only
erased with probability $x$.

\section{Comparison with numerical simulations}
We now present a comparison of the predictions of our RG approach 
with numerical simulations. We first
used a parity check matrix corresponding to a $(3,5)$ regular Gallager code
with $N = 60$ and $k=24$. That is, each of the 36 rows in the parity check
matrix had 5 entries that were ones (the rest were zeros), and each of
the 60 columns had 3 entries which were zeros. There were no hidden nodes. 

We also took care to ensure
that no two parity checks shared more than one variable node. That meant that
there were no loops of length four, so we could use the RG transformation
(\ref{rg_trans_q_noleaf}) (an RG transformation of ``depth'' 2) 
whenever we renormalized away a non-leaf variable
node. We renormalized nodes away until we were left with 7 nodes, and then
finished the computation exactly.

We considered erasure rates $x$ at intervals of $.05$ between $x=0$ and
$x=1$. When we used the RG approximation, we averaged our decoding failure
rates $b_i$ over all 100 nodes $i$ to get an overall bit error rate. Our
numerical simulations consisted of 1000 trials at each erasure rate, decoded
according to the standard BP decoding algorithm.

\begin{figure}[htb]
\centerline{
	\psfig{file=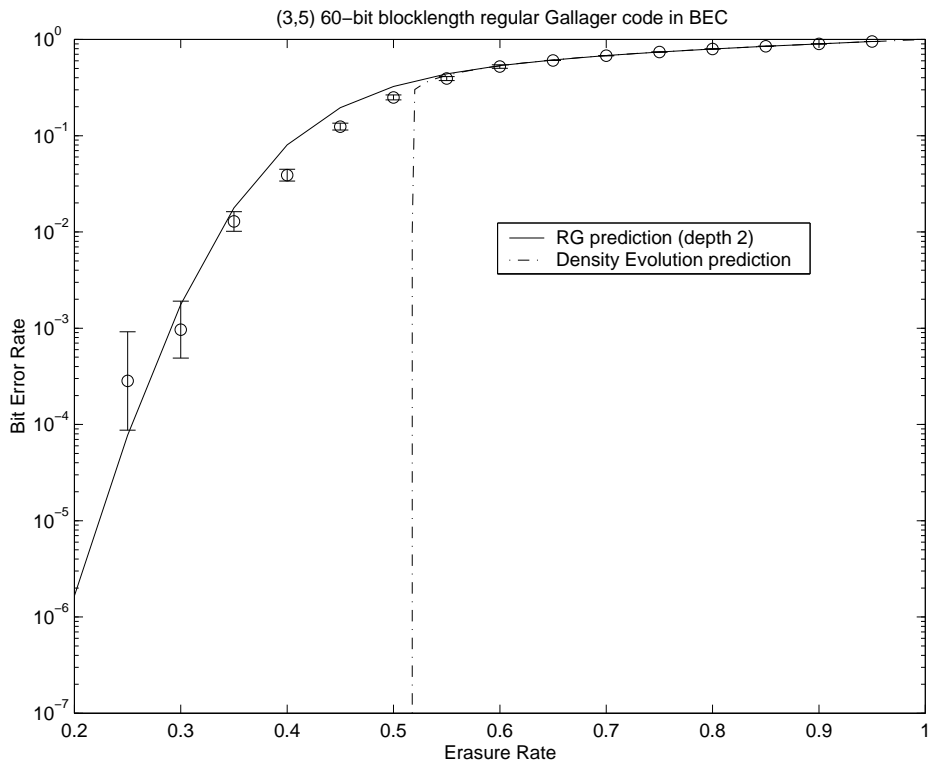,height=4in}}
\caption[]{Simulation results compared with RG and density evolution
predictions for a small rate 2/5 60-bit blocklength regular Gallager
code.}
\label{fig:8}
\end{figure}

Our results are presented in figure \ref{fig:8}, where we compare the simulation results
with the prediction of our RG approach and the density evolution approach.
As one can see, the agreement between the RG approach and simulations is
quite good.

The density evolution prediction is precisely the same as it would be in
the infinite-blocklength limit. Of course, nobody claims that the
density evolution approach should be taken seriously for blocklengths as low
as 60, and figure \ref{fig:8} 
shows why: the density-evolution prediction of
a threshold-like
behaviour is completely incorrect for small or medium blocklength regular
Gallager codes. 

We then constructed, by a somewhat random procedure, a particular
irregular Gallager code of rate $2/5$ and
blocklength $N=100$. Each variable node belonged
to between one and four parity checks, and each parity check involved
between three and five variable nodes. No special effort was made to
construct a particularly good error-correcting code, but 
we did ensure that the Tanner graph
had no short loops of length four or six (counting
both variable and check nodes). 
That meant that all local neighborhoods could be considered 
tree-like up to RG transformations of depth 3. 

Our procedures were the same as described for the regular Gallager code
except that we implemented RG transformations of depth 1, 2, and 3.
Our numerical simulations consisted of 5000 trials for all erasure
rates $x \le .6$, and 1000 trials for higher erasure rates. 

\begin{figure}[htb]
\centerline{
	\psfig{file=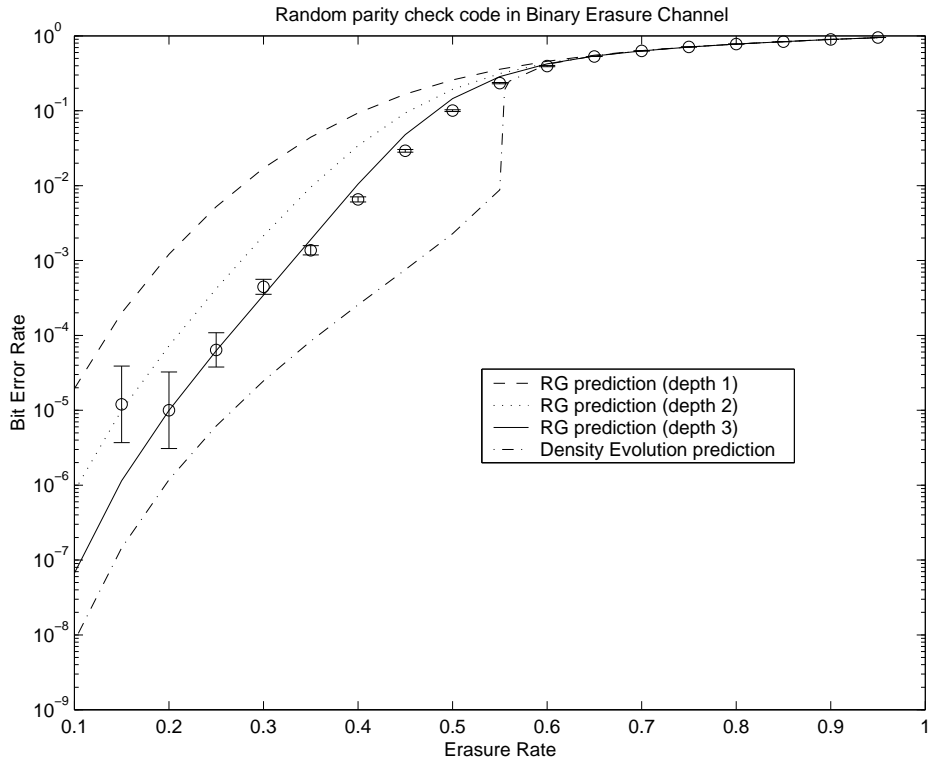,height=4in}}
\caption[]{Simulation results compared with RG predictions using depths
from one to three and density evolution
predictions for a small rate 2/5 100-bit blocklength irregular Gallager
code.}
\label{fig:randplot}
\end{figure}

\begin{figure}[htb]
\centerline{
	\psfig{file=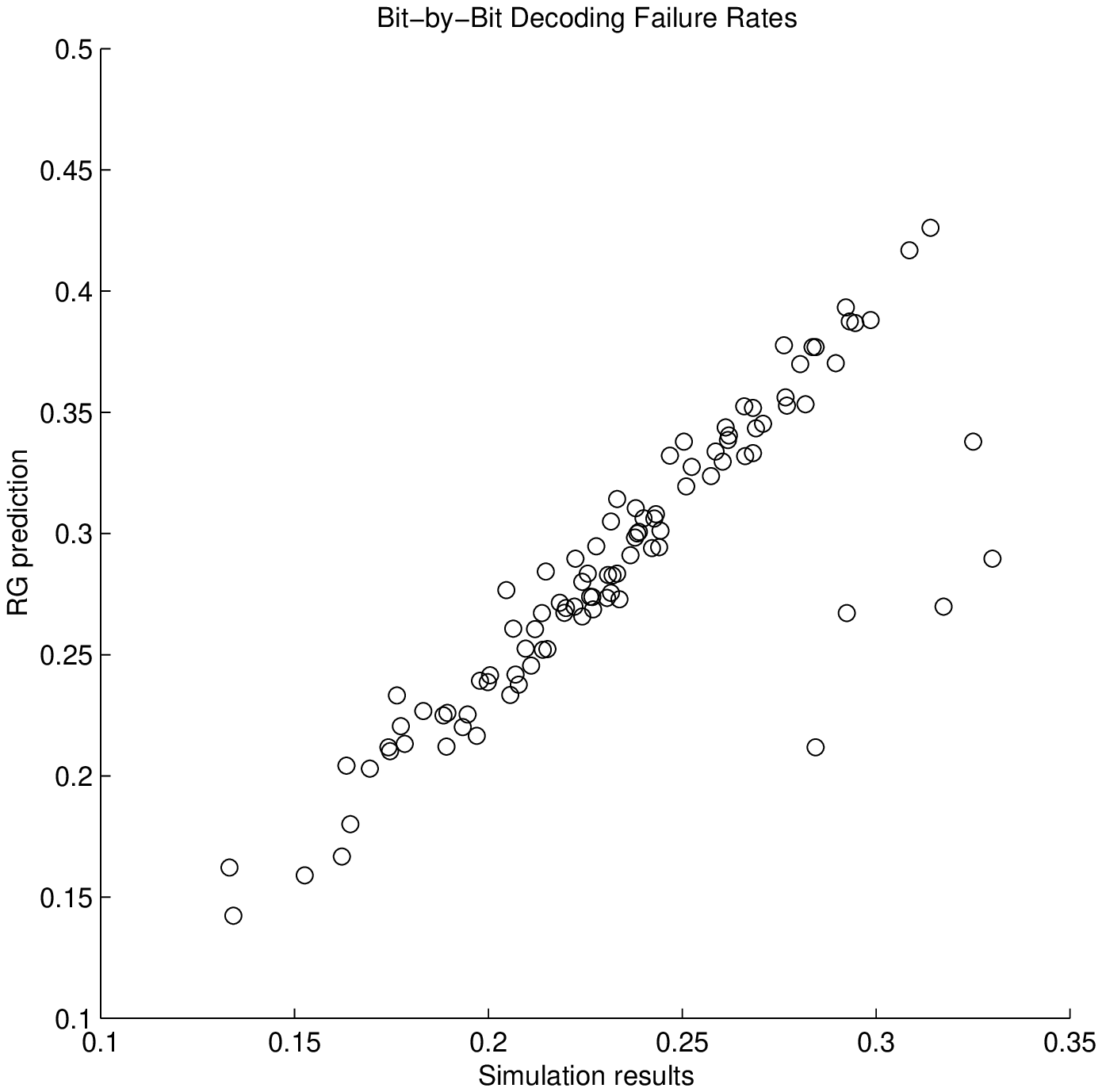,height=4in}}
\caption[]{Bit-by-bit comparison of simulation results 
and RG predictions for a small rate 2/5 100-bit blocklength irregular Gallager
code at an erasure rate of $x=.55$ in the BEC.}
\label{fig:scatter}
\end{figure}

In figure \ref{fig:randplot}, 
we compare the simulation results with the prediction of
our RG approach for the bit error rate averaged over all nodes. As one
can see, the agreement is remarkably good, especially for the RG 
transformations of depth 3. The density evolution prediction spuriously
shows a
quasi-threshold behavior around $x \approx .555$.

The irregular Gallager code has interesting variation in its bit error
rates across the different bits of the code. In figure \ref{fig:scatter}
we plot the predicted (using depth 3 RG transformations) and 
simulated bit error rates for every bit in the code
at an erasure rate of $x=.55$. This plot demonstrates 
that the RG approach can in fact predict the
bit-by-bit variation in the bit error rate. Although the
RG prediction is systematically slightly too high at this erasure
rate, it captures the
ordering of how easily the bits are decoded quite well.

\section{Extension to the Gaussian noise channel.}
\subsection{Background}
In this section, we consider the extension of the RG approach to the
additive white Gaussian noise (AWGN) channel. We will build on the Gaussian
approximation to density evolution for the AWGN channel described
by Chung, et. al. \cite{ChungGauss}, so we first describe that approximation.

In the AWGN channel, there are only two possible inputs, {\tt 0} and {\tt 1},
but the output alphabet is the set of real numbers: if $x$ is the input,
then the output would be $y = (-1)^x + z$, where $z$ is a Gaussian random
variable with zero mean and variance $\sigma^2$. For each received
bit $i$ in the code, we can compute the log-likelihood ratio 
$m_i^0 = \ln(p(y_i|x_i=0)/p(y_i|x_i=1))$ 
which tells us the relative log-likelihood ratio
that the transmitted bit $i$ was a zero given the received real number is 
$y_i$.

We assume that we are
again dealing with codes defined by generalized parity check matrices, that
we always transmit the all-zeros codeword, and
that the decoding algorithm is the sum-product belief propagation algorithm.
In this decoding algorithm, we iteratively solve for real-valued messages:
$m_{ia}$ from variable nodes $i$ to check nodes $a$; and $m_{ai}$ from
check nodes $a$ to variable nodes $i$. The messages $m_{ia}$ are
log-likelihood ratios by which the node $i$ informs the node $a$ of its
probability of being a {\tt 0} or {\tt 1}. For example, 
$m_{ia} \rightarrow \infty$ means
that node $i$ is certain it should be a {\tt 0}, while $m_{ia}=1$ means that
variable node 
$i$ is telling check node $a$ that $\ln(p(x_i= 0)/p(x_i=1)) = 1$. 
The messages $m_{ai}$
are log-likelihood ratios which should be interpreted as information from
the check node $a$ to the variable node $i$ about what state node $i$ should
be in.

In the sum-product algorithm, the messages are
iteratively solved according to the update rules:
\BE
m_{ia} = \sum_{b \in N(i) \backslash a} m_{bi} + m_i^0
\EE
(if $i$ is a hidden node, the $m_i^0$ term is omitted)
and
\BE
\tanh \left(m_{ai}/2 \right) = \prod_{j \in N(a) \backslash i} \tanh 
\left( m_{ja}/2 \right).
\EE

In the density evolution approach for the AWGN channel, one considers the
probability distributions $p(m_{ia})$ and $p(m_{ai})$ for the messages where
the probability distribution is an 
average over all possible received blocks. 
A
distribution $f(x)$ is called {\it consistent} if $f(x) = f(-x)e^x$
for all $x$ \cite{Richardson2001}. Richardson and Urbanke 
\cite{RichardsonUrbanke2001} proved that the consistency condition will be
preserved for the message probability distributions for all messages under
sum-product decoding. If we approximate the probability distributions
$p(m_{ia})$ and $p(m_{ai})$ as Gaussian distributions, the consistency
condition means the means $\mu$ of these distributions will be related to
the variances $\sigma^2$ by $\sigma^2 = 2 \mu$. That means that we can
characterize the message 
probability distributions by a single parameter: their mean.

Thus, by making the approximation that the message probability distributions
are Gaussians, one can reduce the density evolution equations for the
AWGN channel to self-consistent equations for the means $u_{ia}$ of the
probability distributions of messages 
from variable nodes $i$ to check nodes $a$, and the
means $v_{ai}$ of the probability distributions of messages from
check nodes $a$ to variable nodes $i$. These equations are
\BE
\label{de_awgn_v}
v_{ia} = u_i^0 + \sum_{b \in N(i) \backslash a} u_{bi}
\EE
where $u^0$ is the mean value of $m_i^0$ (this term is omitted for hidden
nodes),
and
\BE
\label{de_awgn_u}
\phi(u_{ai}) = 1 - \prod_{j \in N(a) \backslash i} (1- \phi(v_{ja}))
\EE
where $\phi(x)$ is a function defined by
\BE
\phi(x) \equiv 1 - \frac {1}{\sqrt{4 \pi x}} \int_{- \infty}^{\infty}
\tanh{\frac{u}{2}} e^{-\frac {(u-x)^2}{4x}} du
\EE

$\phi(x)$ can be approximated in a form that reproduces the correct
limits as $x \rightarrow 0$ and $x \rightarrow \infty$ and
is more convenient for numerical
purposes. We choose
\BE
\phi(x) \approx
\frac{e^{-x/4}}{\sqrt{1+\beta x}}
\left[1+(\sqrt{\beta \pi -1}) \frac{\alpha x}{1+\alpha x} \right]
\EE
This form automatically 
has the correct leading behavior as $x \rightarrow 0$ and 
$x \rightarrow \infty$ for any $\alpha$ and $\beta$. We fix $\alpha$
and $\beta$ by matching the leading corrections in the two limits. We
find $\alpha \approx 0.163489$ and $\beta \approx 0.634765$. This 
approximation to $\phi(x)$ is quite good for all values of $x$.

\subsection{RG transformations for the AWGN channel}
The density evolution equations (\ref{de_awgn_v}) and (\ref{de_awgn_u}) for the
AWGN channel under the Gaussian approximation are analogs of the density
evolution equations (\ref{q_de_general}) and (\ref{p_de_general}) for the
BEC channel. Our RG procedure for the AWGN channel will be almost exactly
the same as for the BEC channel; the main difference is that we need to
change the RG transformations. 

Just as before, we can construct a set of RG transformations
which exactly reproduce the density evolution equations for a tree-like 
graph. We create a decorated Tanner/Wiberg graph for the code by keeping 
$u_{ai}$ and $v_{ia}$ variables between each pair of connected nodes. The
$u_{ai}$ variables are initialized to $\infty$, while the $v_{ia}$ variables
are initialized to $u^0$, unless the $i$th node is a hidden node, in which
case the $v_{ia}$ are initialized to zero. We also introduce the variables
$h_i$ (analogous to $b_i$ in the BEC) which are initialized like the $v_{ia}$
variables.

If we renormalize away a leaf check node $a$ attached to a check node $i$,
we find the other check nodes $b$ attached to $i$ 
and apply the RG transformations
\BE
\label{rg_trans_awgn_v}
v_{ib} \leftarrow v_{ib} + u_{ai}
\EE
and
\BE
\label{rg_trans_awgn_h}
h_i \leftarrow h_i + u_{ai}
\EE
while if we renormalize away a leaf variable node $i$ attached to a check
node $a$, we find the other variable nodes $j$ attached to $a$ and
apply the RG transformation
\BE
\label{rg_trans_awgn_u}
u_{aj} \leftarrow \phi^{-1} \left(1-(1-\phi(u_{aj}))(1-\phi(v_{ia})) \right)
\EE
Note that with each renormalization of $v_{ib}$, the magnitude of $v_{ib}$ will
increase, while with each renormalization of $u_{aj}$, the magnitude of
$u_{aj}$ will decrease.

When we renormalize away a non-leaf variable node $i$ which is attached to
check nodes $a$, $b$, etc., we need to renormalize the variables like
$u_{aj}$, where $j$ is another variable node attached to check node $a$.
Just as for the BEC, we should consider a local neighborhood of nodes
around the node $i$. For example, if no variable nodes $j$ share two
check nodes with $i$ (there are no local loops of length four)
then we can use the depth two RG transformation
\BE
\label{rg_trans_awgn_u_nonleaf}
u_{aj} \leftarrow \phi^{-1} \left(1-(1-\phi(u_{aj}))(1-\phi(v_{ia}^{\rm eff})
) \right)
\EE
where 
\BE
v_{ia}^{\rm eff} = v_{ia} + \sum_{b \in N(i) \backslash a} 
\phi^{-1} \left( 1 - (1-\phi(u_{bi})) \prod_{k \in N(b) \backslash i}
(1-\phi(v_{kb})) \right)
\EE

The RG procedure proceeds as in the BEC case until the final computation of the
bit error rate. For the AWGN channel, it will not normally be convenient to
stop the RG procedure before renormalizing all the way down to the ``target''
node, because it is not simple to do an exact computation even with just a
few nodes in the code. When we have renormalized all but our target node $i$,
we will be left with a final renormalized value of $h_i$. Our Gaussian 
approximation tells us that the probability distribution for the 
node $i$ being decoded
as a zero will be a Gaussian with mean $h_i$ and variance $2 h_i$. 
Decoding failures correspond to those parts of the probability distribution
which are below zero. Thus, our theoretical prediction for
the bit error rate at node $i$ will be
\BE
b_i = \frac{1}{\sqrt{8 \pi h_i}} 
\int_{- \infty}^0 e^{-\frac{(x-h_i)^2}{4 h_i}} dx.
\EE

\section{Speculations on the design of codes}
Given that the density evolution method has been used as a guide to designing
the best-known practical codes, it is natural to expect that we could design
even better codes using the RG approach. With the RG approach, we can
input a code defined by an arbitrary generalized parity check matrix, and
obtain as output a prediction of the bit error rate at each node. We could
use this output as the objective function for a guided search through the
space of possible codes. For example, say that we would like to find a
$N=100$ rate $1/2$ code with no hidden states 
that achieves a bit error rate of less than $10^{-4}$
at the smallest possible signal-to-noise ratio for the AWGN channel. We 
could repeatedly evaluate codes using the RG approach, and use any 
available search
technique (greedy descent, simulated annealing, genetic algorithms, etc.)
to search through the space of valid parity check matrices.
Because we can directly focus on the correct figure of merit (the bit error
rate itself, rather than the threshold in the infinite blocklength limit), one
expects the search to improve on the results obtained using density evolution.

A couple of comments are in order. First, because we have information
about the bit error rate at every node (see figure \ref{fig:scatter}), 
we might be able 
to use that information to
guide the search. For example, it might make sense to ``strengthen'' a
variable node with a high bit error rate by adding it to more parity checks,
or one could choose to ``weaken'' nodes with a low bit error rate by turning
them into hidden nodes (thus increasing the rate).

On the other hand, computing the bit error rate of every node will obviously
slow down a search. It may be worthwhile, at least for large
blocklengths, to restrict
oneself to those codes for which there are only a small number of different
classes of nodes (defined in terms of the local neighborhoods of the nodes).
Most of the best-known codes are of this type. Rather
than computing the bit error rate for every variable node, one could then
compute the bit error rate for just one representative of each class of 
variable node. For example, for a regular Gallager code, each node has the
same local neighborhood, so any node can be chosen as a representative of all
the nodes. The error made in this approach can be estimated by comparing
bit error rates of different nodes of the same class. For actual 
finite-sized regular Gallager
codes, we find that the RG approach will give very similar predictions for
each of the nodes, so that the error made by just considering a single
variable node as a representative of all of them is quite small.

\section*{Acknowledgements}
We thank Bill Freeman, Yair Weiss, Michael Mitzenmacher, Dave Forney, and
David MacKay for helpful discussions. Jean-Philippe Bouchaud thanks
Harvard University and Daniel Fisher for their hospitality during the period
when this work was done.

\end{document}